\documentclass[letterpaper]{article} 
\usepackage[preprint]{aaai2027}  
\usepackage[hyphens]{url}  
\usepackage{graphicx} 
\usepackage{natbib}  
\usepackage{caption} 
\usepackage{amsmath,amssymb}
\usepackage{algorithm}
\usepackage{algorithmic}

\usepackage{newfloat}
\usepackage{listings}
\DeclareCaptionStyle{ruled}{labelfont=normalfont,labelsep=colon,strut=off} 
\floatstyle{ruled}
\newfloat{listing}{tb}{lst}{}
\floatname{listing}{Listing}

\usepackage{booktabs}
\usepackage[most]{tcolorbox}

\title{ACE-Cap: Active Evidence Acquisition via Agentic Co-Evolution for Long-Paragraph Fine-Grained Audio Captioning}
\author{
    Fengji Ma\textsuperscript{\rm 1,\rm 2}\thanks{Work done during an internship at KlingAI Research.},
    Yan Rong\textsuperscript{\rm 1,\rm 2}\footnotemark[1],
    Xu Li\textsuperscript{\rm 2}\thanks{Project Leader.},
    Xuenan Xu\textsuperscript{\rm 2},
    Chen Zhang\textsuperscript{\rm 2},
    Li Liu\textsuperscript{\rm 1}\corresponding
}
\affiliations{
    \textsuperscript{\rm 1}The Hong Kong University of Science and Technology (Guangzhou)\\
    \textsuperscript{\rm 2}KlingAI Research
}

\begin{document}

\maketitle

\begin{abstract}
Long-paragraph fine-grained audio captioning requires models to recover diverse acoustic facts while avoiding omissions and unsupported details. However, prevailing captioners remain passive one-shot generators: once a relevant detail is overlooked, they cannot identify the evidence gap, query the audio for targeted information, or adaptively determine when sufficient evidence has been collected. We formulate long-paragraph fine-grained audio captioning as an active evidence-acquisition problem and introduce Agentic Co-Evolution for Captioning (ACE-Cap). The framework uses multi-turn interaction between a Composer and an Instruct model to form a closed evidence-acquisition loop. A Captioner first produces an initial description. Conditioned only on this description and the interaction history, a text-only Composer asks targeted questions about unresolved acoustic attributes, while an audio-conditioned Instruct model provides grounded answers. The Composer then decides when to terminate the interaction and synthesizes the accumulated evidence into a final caption. ACE-Cap trains these roles through a unified gold-to-prediction reward derived from fixed, gold-grounded multiple-choice questions and a frozen caption-only judge. To assign credit within variable-length interactions, LOOP-GRPO replaces the trajectory-wide scalar advantage with span-aligned signals: leave-one-out contributions computed from the accumulated evidence for individual questions, a quality--cost utility for stopping, and an evidence-preservation utility for final synthesis. Role-wise warm-up followed by alternating Composer and Instruct optimization keeps each update a well-defined single-policy problem while allowing the two roles to co-evolve. Together, ACE-Cap turns captioning from passive one-shot generation into an adaptive process that learns what evidence to acquire, when to stop acquiring it, and how to preserve it in a long-paragraph fine-grained audio caption.
\end{abstract}


\begin{figure*}[t!]
    \centering
    \includegraphics[width=0.83\textwidth]{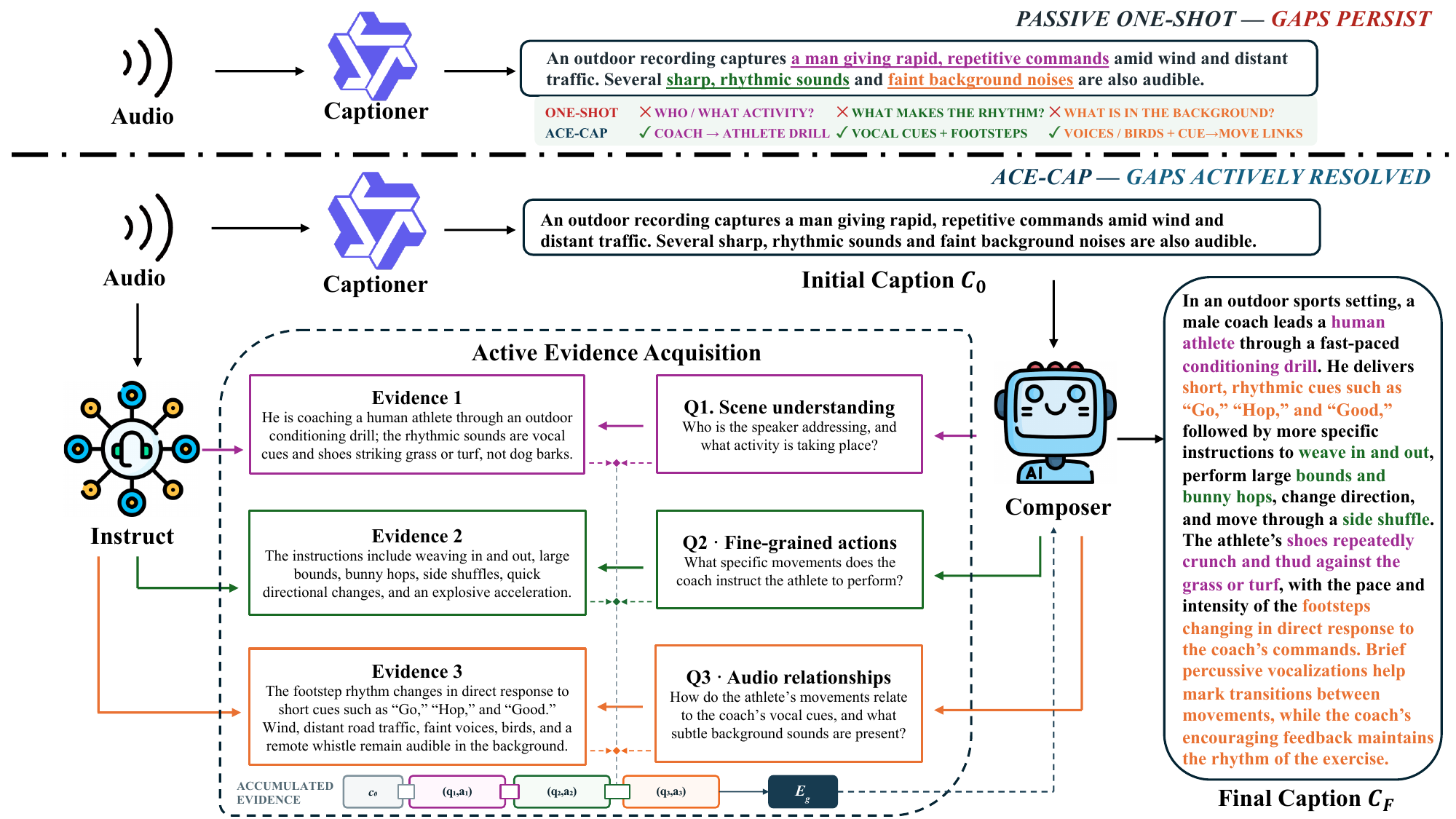}
    \caption{Paradigm comparison for long-paragraph fine-grained audio captioning. Passive one-shot captioners produce a caption without actively resolving missing details (top). ACE-Cap instead uses a text-only Composer to query an audio-conditioned Instruct model, accumulates the resulting question--answer evidence as Accumulated Evidence $E_g$, and synthesizes a detailed final caption (bottom).}
    \label{fig:teaser}
\end{figure*}

\section{Introduction}


Automated audio captioning (AAC) translates an acoustic scene into a natural-language description.
Early benchmarks such as AudioCaps and Clotho enabled AAC research at scale by providing standardized benchmarks.
Advances in Transformer-based audio-language architectures and large-scale fine-grained datasets have substantially improved the fluency and accuracy of audio captioning \citep{audio_captioning_transformer,audio_flamingo2,fusionaudio,audiomap}.
The emerging goal, however, is not merely to produce a longer sentence.
A useful fine-grained caption must recover a broad set of acoustic facts, including speech, music, sound events, ambience, recording characteristics, and temporal relations.
Meeting this broader target exposes a basic limitation of the prevailing captioning framework.

Most captioners remain \emph{passive one-shot generators}: they encode the audio and commit to a complete description in a single decoding process. Once a relevant detail is overlooked, the model has no explicit mechanism to diagnose what is missing, revisit the audio with a targeted query, or allocate additional computation to an uncertain acoustic dimension. Increasing model or data scale improves the underlying representation, but does not by itself provide an active information-acquisition procedure. Consequently, vague or omitted acoustic details can persist into the final caption without an opportunity for targeted correction.

A natural response is to replace single-pass generation with \emph{active evidence acquisition}, as illustrated in Fig.~\ref{fig:teaser}. An initial caption serves as a provisional hypothesis rather than an irrevocable output. A text-only questioner inspects this hypothesis and the interaction history to identify an unresolved acoustic dimension, while an audio-conditioned answerer revisits the recording to provide targeted evidence. Repeated question--answer turns form the \textbf{Accumulated Evidence} ($E_g$); the system then decides when the evidence is sufficient and synthesizes it into a final caption. This formulation provides the missing correction mechanism, but training it introduces two additional problems that do not arise in the same form for one-shot captioning.

The first training problem is \emph{credit assignment} across heterogeneous actions. Existing caption metrics primarily score a completed output, from learned semantic similarity to multi-factor and graph-grounded evaluation \citep{fense,xace,graphscore}. Likewise, standard Group Relative Policy Optimization (GRPO) derives one group-relative advantage from a completion-level reward and applies it across the sampled completion \citep{deepseekmath}. In an active captioning trajectory, however, one question may reveal a unique event, another may repeat known information, a stopping action may be premature or wastefully late, and the final synthesis may discard evidence that was successfully collected. Rewarding each turn by its score improvement over the immediately preceding state is also insufficient: this sequential difference depends on the order-specific history, gives early questions first-mover credit, and can undervalue later but more precise questions after the score has partially saturated. What is needed instead is post-rollout counterfactual credit: after the interaction ends, evaluate how much the accumulated evidence state degrades when one question--answer turn is removed. This leave-one-out view follows counterfactual credit assignment in cooperative reinforcement learning \citep{coma}, but adapts it to questioning, stopping, and synthesis in active captioning.

The second training problem is instability caused by jointly adapting the interacting roles. As the questioner changes, the distribution of questions presented to the answerer also changes; as the answerer changes, the responses observed by the questioner change in turn. Each role is therefore optimized against a moving target, a known source of training instability in multi-agent learning \citep{marl_nonstationarity}. Sequential or multi-timescale updates can mitigate this instability \citep{multitimescale_marl}, yet alternating updates alone are insufficient when initially weak roles generate poor supervision for one another. Training therefore requires both role-specific preparation and a controlled schedule in which only one policy changes at a time.

To realize this active captioning process while addressing its two training problems, we propose \textbf{Agentic Co-Evolution for Captioning (ACE-Cap)}, a framework for fine-grained audio captioning. A Captioner first produces the provisional caption. A text-only Composer then asks what remains unresolved, an audio-conditioned Instruct model listens to the audio and answers, and the Composer decides when to stop and synthesizes the Accumulated Evidence. A unified gold-to-prediction (G2P) reward evaluates every textual state through fixed multiple-choice questions constructed offline from each audio--gold-caption pair and a frozen caption-only judge that sees only the candidate text. On top of G2P, our \textbf{LOOP-GRPO} (Leave-One-Out Per-turn GRPO) assigns each question its leave-one-out contribution to the accumulated evidence, trains stopping according to a quality--cost trade-off, and rewards the final caption for preserving acquired evidence. Finally, the three roles receive separate supervised and GRPO warm-ups; the Captioner is then fixed while Composer and Instruct are optimized alternately, so every agentic update remains a well-defined single-policy problem.

Our contributions are threefold:
\begin{itemize}
    \item We formulate fine-grained audio captioning as active evidence acquisition and introduce ACE-Cap, which decomposes generation into initial captioning, targeted questioning, audio-grounded answering, adaptive termination, and evidence-aware summarization.
    \item We introduce LOOP-GRPO under a unified G2P reward, providing action-aligned credit for what to ask, when to stop, and how to preserve the acquired information; unlike adjacent reward differences, its question credit measures each turn's contribution to the accumulated evidence state and is therefore more robust to turn order and score saturation.
    \item We develop a warm-started co-evolution strategy that separately prepares all three roles, freezes the Captioner during agentic optimization, and alternates updates between Composer and Instruct so that each GRPO stage remains a well-defined single-policy problem.
\end{itemize}

\section{Related Works}

\subsection{Audio Captioning and Agentic Interaction}

Audio captioning developed through AudioCaps and Clotho
\citep{audiocaps,Clotho}, transformer encoder--decoders
\citep{audio_captioning_transformer}, and weakly labeled pretraining in
WavCaps \citep{wavcaps}. General audio--language models include Pengi
\citep{pengi}, Audio Flamingo \citep{audio_flamingo}, Audio Flamingo 2
\citep{audio_flamingo2}, and FusionAudio \citep{fusionaudio}. Agentic systems
couple reasoning with actions \citep{react}, coordinate perceptual experts for
multimodality-to-multiaudio generation \citep{mm_react,raudiogenie}, support
audio dialogue and immersive audiobook generation
\citep{audio_flamingo,dopamine}, or acquire and iteratively refine acoustic
evidence for reasoning
\citep{omniagent_active_perception,audiogenie_reasoner}. These systems broaden
acoustic coverage and interaction, but existing audio agents primarily target
downstream reasoning or audio-content generation, while captioners still
produce descriptions in one pass. ACE-Cap instead uses interaction to identify
and repair missing acoustic evidence before synthesizing a long-form caption.

\subsection{Reinforcement Learning for Interactive Audio Captioning}

Captioning has been optimized with self-critical training \citep{scst} and
audio-specific reinforcement learning \citep{crnn_gru_rl_aac}, supported by
semantic, LLM-based, and factorized caption metrics
\citep{fense,clair_a,xace}. For multi-turn policies, GRPO supplies
group-relative optimization \citep{deepseekmath}, turn-level methods assign
intermediate credit \citep{multiturn_credit_assignment}, and counterfactual
learning estimates action contributions \citep{coma}. Adaptive agents
additionally face non-stationarity \citep{marl_nonstationarity}, which
multi-timescale updates can mitigate \citep{multitimescale_marl}. ACE-Cap
combines these directions through a gold-derived G2P reward evaluated by a
frozen caption-only judge, LOOP-GRPO credits for questioning, stopping, and
synthesis, and alternating Composer--Instruct optimization. This design
targets evidence-preserving caption generation rather than answer generation
alone.

\section{Methodology}

\subsection{ACE-Cap Framework}

Let $\mathcal{D}=\{(x_n,c_n^\star)\}_{n=1}^{|\mathcal{D}|}$ contain audio
$x_n$ and its gold long-form caption $c_n^\star$; we omit $n$ below.
ACE-Cap has three roles as shown in Fig.~\ref{fig:teaser}. The audio-conditioned Captioner $f_\phi$ produces
$c_0=f_\phi(x)$; the text-only Composer $\pi_\theta$ asks questions, stops,
and synthesizes the final caption; and the audio-conditioned Instruct model
$p_\psi$ answers those questions. Because the Composer never receives audio,
missing acoustic evidence can enter only through Instruct answers.

Offline Gold-MCQs derived from each audio--caption pair define a shared
reward. A frozen caption-only judge scores candidate text without seeing the
audio or gold caption. These supervision artifacts are used only during
training and evaluation.

\subsection{Unified Gold-to-Prediction Reward}

\begin{figure*}[t!]
\centering
\includegraphics[width=0.85\textwidth]{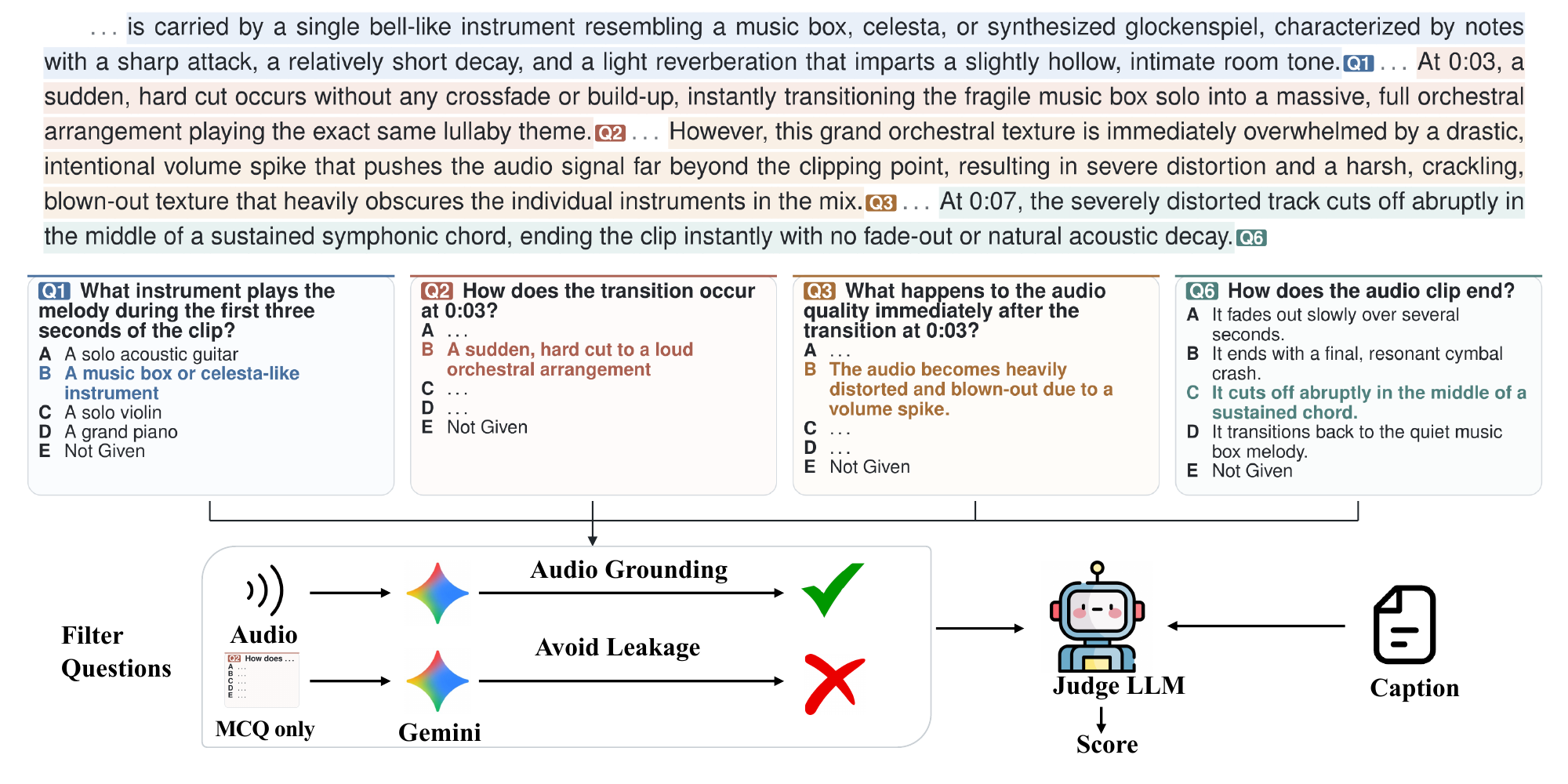}
\caption{Gold-MCQ construction and G2P scoring. Colors link fixed MCQs to
gold-caption evidence spans. Candidate questions are filtered offline; during
GRPO, the frozen judge scores only the candidate caption against the retained
MCQs, without audio or the gold caption.}
\label{fig:g2p-bank-example}
\end{figure*}

For a pair $(x,c^\star)$, the offline stage constructs a fixed question bank
\begin{equation}
\mathcal{M}^\star(x,c^\star)
=\{(m_j,o_j^\star)\}_{j=1}^{L},
\label{eq:gold-mcq-bank}
\end{equation}
where $m_j$ contains one question and five options, and $o_j^\star\in\{\mathrm{A},\mathrm{B},\mathrm{C},\mathrm{D}\}$ is the correct option. Option $\mathrm{E}$ is always \textit{Not Given} and is never a gold answer. Given a candidate text $y$, the frozen judge $\mathcal{J}$ predicts $\hat{o}_j(y)=\mathcal{J}(y,m_j)$. Following~\cite{audiomap}, we assign the graded per-question score
\begin{equation}
s_j(y)=
\begin{cases}
1, & \hat{o}_j(y)=o_j^\star,\\
-0.5, & \hat{o}_j(y)=\mathrm{E},\\
-1, & \text{otherwise},
\end{cases}
\label{eq:g2p-item-score}
\end{equation}
and define the gold-to-prediction (G2P) reward as
\begin{equation}
R_G(y)=\frac{1}{L}\sum_{j=1}^{L}s_j(y).
\label{eq:g2p-reward}
\end{equation}
Thus, supported, omitted, and conflicting facts receive $1$, $-0.5$, and
$-1$, respectively. All stages use Eq.~\eqref{eq:g2p-reward} with the same
judge and scale. Fig.~\ref{fig:g2p-bank-example} shows the localized
grounding. Because the bank is fixed, $R_G$ evaluates only represented facts
and cannot detect unsupported claims outside it.

\subsection{Agentic Captioning Rollout}

For each audio, GRPO samples a group of $G$ Composer rollouts indexed by $g\in\{1,\ldots,G\}$. Before its $t$-th possible question, the Composer observes the text history
\begin{equation}
h_{g,t}=\left(c_0,(q_{g,s},a_{g,s})_{s=1}^{t-1}\right)
\label{eq:composer-history}
\end{equation}
and samples an action
\begin{equation}
z_{g,t}\sim\pi_\theta(\cdot\mid h_{g,t}),\qquad
z_{g,t}\in\{\operatorname{ASK}(q_{g,t}),\operatorname{READY}\}.
\label{eq:composer-action}
\end{equation}
For an $\operatorname{ASK}$ action, the Instruct model returns
\begin{equation}
a_{g,t}\sim p_\psi(\cdot\mid x,h_{g,t},q_{g,t}).
\label{eq:instruct-answer}
\end{equation}
The answer is a loss-masked environment observation. If
$\operatorname{READY}$ precedes question $t$, then $T_g=t-1$; otherwise, the
scheduler forces finalization at $T_g=K$ without creating a policy stop token.

After the interaction stops, the accumulated evidence state, hereafter \textbf{Accumulated Evidence} ($E_g$), and its leave-one-out counterpart are
\begin{align}
E_g
&=\left(c_0,(q_{g,t},a_{g,t})_{t=1}^{T_g}\right),
\label{eq:accumulated-evidence}\\
E_g^{-t}
&=\left(c_0,(q_{g,s},a_{g,s})_{\substack{1\leq s\leq T_g\\s\ne t}}\right).
\label{eq:loo-evidence}
\end{align}
The Composer then generates the final caption from
Eq.~\eqref{eq:accumulated-evidence},
\begin{equation}
c_g^F\sim\pi_\theta(\cdot\mid E_g,\operatorname{FINALIZE}).
\label{eq:final-caption-generation}
\end{equation}
We use only three score types throughout the Composer objective:
\begin{equation}
r^0=R_G(c_0),\qquad
r_g^E=R_G(\operatorname{ser}(E_g)),\qquad
r_g^F=R_G(c_g^F).
\label{eq:three-scores}
\end{equation}
Here, $\operatorname{ser}(E_g)$ is the judge input formed by deterministically
concatenating $c_0$ and all completed question--answer pairs in
Eq.~\eqref{eq:accumulated-evidence}, in interaction order and with role
delimiters. Questions are marked non-assertive, so only $c_0$ and answers
count as evidence.

\begin{algorithm}[t]
\caption{ACE-Cap Training}
\label{alg:ace-cap}
\begin{algorithmic}[1]
\REQUIRE $\mathcal{D}$, offline banks $\{\mathcal{M}_n^\star\}$, rollout size $G$, budget $K$
\STATE SFT Captioner $f_\phi$, Instruct $p_\psi$, and Composer $\pi_\theta$
\STATE Warm up $f_\phi$ by GRPO using Eq.~\eqref{eq:captioner-reward}
\STATE Warm up $p_\psi$ by per-QA GRPO using Eq.~\eqref{eq:instruct-reward}
\STATE Warm up $\pi_\theta$ by LOOP-GRPO with fixed $p_\psi$
\STATE Freeze $f_\phi$ permanently
\FOR{each co-evolution cycle}
    \STATE Freeze $\pi_\theta$; harvest prompts and update $p_\psi$
    \STATE Freeze $p_\psi$; roll out $G$ trajectories and update $\pi_\theta$
\ENDFOR
\RETURN $f_\phi$, $p_\psi$, and $\pi_\theta$
\end{algorithmic}
\end{algorithm}

\subsection{LOOP-GRPO}

Standard GRPO broadcasts one group-relative completion advantage to all
tokens \citep{deepseekmath}. LOOP-GRPO instead assigns separate utilities to
question, stop, and final-caption spans.

\paragraph{Leave-one-out question credit.}
For each question that actually occurs, we first score its leave-one-out evidence state,
\begin{equation}
r_{g,-t}^E=R_G(\operatorname{ser}(E_g^{-t})),
\label{eq:loo-score}
\end{equation}
and assign the raw question utility
\begin{equation}
u_{g,t}^Q=r_g^E-r_{g,-t}^E-c_Q,
\label{eq:question-utility}
\end{equation}
where $c_Q\geq0$ is a per-question cost. Utility is low when removing a
question--answer pair leaves the completed evidence unchanged.

A sequential alternative would use the adjacent difference
\begin{equation}
d_{g,t}
=R_G(\operatorname{ser}(E_{g,t}))
-R_G(\operatorname{ser}(E_{g,t-1})),
\label{eq:sequential-delta}
\end{equation}
where $E_{g,t}$ is the prefix after turn $t$. This marginal depends on facts
already exposed and can favor early coarse questions under a saturating
score. Equation~\eqref{eq:question-utility} instead removes each turn from
the same final evidence state, following counterfactual credit assignment
\citep{coma}. Serialization and judge behavior may still introduce residual
order sensitivity.

\paragraph{Stop and final-caption credits.}
The stopping and synthesis actions depend jointly on the final outcome and on whether collected evidence survives summarization. We define the evidence-preserving base utility
\begin{equation}
b_g=(r_g^F-r^0)
+\lambda\left[\min(r_g^E,r_g^F)-r^0\right],
\qquad \lambda\geq0.
\label{eq:base-utility}
\end{equation}
The first term measures improvement over $c_0$; the bottleneck term is high
only when both collected evidence and its synthesis preserve the same facts.
For a sampled $\operatorname{READY}$ action, the stop utility is
\begin{equation}
u_g^S=b_g-c_ST_g,
\label{eq:stop-utility}
\end{equation}
where $c_S\geq0$ trades quality against interaction length. The final-caption
utility is
\begin{equation}
u_g^F=b_g.
\label{eq:final-utility}
\end{equation}
The costs in Eqs.~\eqref{eq:question-utility} and
\eqref{eq:stop-utility} control individual questions and trajectory length,
respectively. For budget truncation, Eq.~\eqref{eq:stop-utility} is omitted,
while questions and synthesis remain trained through
Eqs.~\eqref{eq:question-utility} and \eqref{eq:final-utility}.

\paragraph{Span-aligned group-relative optimization.}
Let $\tau\in\{Q,S,F\}$ denote an action type and let $\Xi_\tau$ collect all valid actions of that type from the $G$ rollouts of the same audio. For an action instance $\xi\in\Xi_\tau$ with raw utility $u(\xi)$, its normalized advantage is
\begin{equation}
A(\xi)=\frac{u(\xi)-\mu_\tau}
{\sigma_\tau+\varepsilon_{\mathrm{norm}}},
\label{eq:type-normalization}
\end{equation}
where $\mu_\tau$ and $\sigma_\tau$ are the mean and standard deviation within
$\Xi_\tau$, preventing scale differences across action types.

Let $\mathcal{I}_{g,t}^Q$, $\mathcal{I}_g^S$, and $\mathcal{I}_g^F$ be the Composer-token spans for the $t$-th question, a sampled $\operatorname{READY}$ action, and the final caption. The token-aligned advantage is
\begin{equation}
A_{g,k}^{\mathrm{tok}}=
\begin{cases}
A_{g,t}^{Q}, & k\in\mathcal{I}_{g,t}^{Q},\\
A_g^{S}, & k\in\mathcal{I}_g^{S},\\
A_g^{F}, & k\in\mathcal{I}_g^{F},\\
0, & \text{otherwise}.
\end{cases}
\label{eq:token-advantage}
\end{equation}
For forced finalization, $\mathcal{I}_g^S=\varnothing$. The Composer loss
mask $M_{g,k}$ also excludes Instruct answers, prompts, user messages,
scheduler tokens, and padding.

For Composer token $y_{g,k}$ with text context $\mathcal{H}_{g,k}$, define
\begin{equation}
\rho_{g,k}(\theta)
=\frac{\pi_\theta(y_{g,k}\mid\mathcal{H}_{g,k})}
{\pi_{\theta_{\mathrm{old}}}(y_{g,k}\mid\mathcal{H}_{g,k})}.
\label{eq:policy-ratio}
\end{equation}
We denote its clipped value by
\begin{equation}
\bar{\rho}_{g,k}(\theta)
=\operatorname{clip}\!\left(
\rho_{g,k}(\theta),
1-\varepsilon_{\mathrm{clip}},
1+\varepsilon_{\mathrm{clip}}
\right).
\label{eq:clipped-ratio}
\end{equation}
LOOP-GRPO preserves the clipped GRPO objective and replaces only its broadcast scalar advantage:
\begin{equation}
\begin{aligned}
\mathcal{L}_{\mathrm{LOOP}}(\theta)
=&-\frac{\sum_{g,k}M_{g,k}
\min\!\left(
\rho_{g,k}A_{g,k}^{\mathrm{tok}},
\bar{\rho}_{g,k}A_{g,k}^{\mathrm{tok}}
\right)}
{\sum_{g,k}M_{g,k}}\\
&+\beta\overline{D}_{\mathrm{KL}}
(\pi_\theta\|\pi_{\mathrm{ref}}),
\end{aligned}
\label{eq:loop-objective}
\end{equation}
where $\overline{D}_{\mathrm{KL}}$ is the masked token-average KL penalty. We use separate constants $\varepsilon_{\mathrm{norm}}$ and $\varepsilon_{\mathrm{clip}}$ for advantage normalization and policy clipping.

\subsection{Role-Wise Warm-Up and Alternating Co-Evolution}

Each role is first trained with supervised fine-tuning on its own interface,
then warmed up by single-policy GRPO while the other components remain fixed.

Captioner GRPO directly uses
\begin{equation}
u^{\mathrm{Cap}}=R_G(c_0).
\label{eq:captioner-reward}
\end{equation}
For Instruct GRPO, a fixed $(x,h_t,q_t)$ is replayed with $G$ answers. Let
$y_t^-$ denote the evidence before an answer and $y_{g,t}^+$ the same context
with sampled answer $a_{g,t}$. Its utility is
\begin{equation}
u_{g,t}^{I}=R_G(y_{g,t}^+)-R_G(y_t^-).
\label{eq:instruct-reward}
\end{equation}
Only answer tokens receive advantage. Composer warm-up applies LOOP-GRPO
with a fixed Instruct model.

After warm-up, the Captioner remains frozen. Instruct phases freeze the
Composer and update $\psi$ on its harvested prompts using
Eq.~\eqref{eq:instruct-reward}; Composer phases freeze Instruct and update
$\theta$ using Eq.~\eqref{eq:loop-objective}. Alternation keeps each update
single-policy while the roles co-evolve through changing prompts and
responses. At inference, all roles are frozen and interact until
$\operatorname{READY}$ or budget $K$; the gold caption, MCQs, and judge are
unused.

\begin{table}[t]
\centering
\footnotesize
\setlength{\tabcolsep}{3.2pt}
\begin{tabular}{@{}lccc@{}}
\toprule
\textbf{Method} & \textbf{MMAU} $\uparrow$ & \textbf{MMAR} $\uparrow$ & \textbf{MMSU} $\uparrow$ \\
\midrule
\multicolumn{4}{@{}l}{\textit{\textbf{Proprietary models}}} \\
GPT-4o Audio           & 62.4 & 59.3 & 56.4 \\
Gemini2.0-Flash       & 58.6 & 50.6 & 51.0 \\
Gemini2.5-Flash       & 65.6 & 58.2 & 58.1 \\
Gemini2.5-Pro         & 70.0 & 64.1 & 71.8 \\
Gemini3.1-Pro         & 71.3 & 65.7 & 69.5 \\
Qwen3.5-Omni-Plus     & 71.6 & 62.6 & 69.8 \\
\midrule
\multicolumn{4}{@{}l}{\textit{\textbf{Open-source models}}} \\
Audio-Flamingo 3-7B     & 47.1 & 36.7 & 42.5 \\
Kimi-Audio-7B            & 54.5 & 39.8 & 43.4 \\
Step-Audio-2-mini-8B   & 62.5 & 44.6 & 53.6 \\
MiDashengLM-7B         & 67.6 & 51.7 & 62.2 \\
Qwen2-Audio-7B         & 63.3 & 44.2 & 53.3 \\
Qwen2.5-Omni-7B        & 65.2 & 51.8 & 60.6 \\
Qwen3-Omni-30B-A3B-Instruct    & \underline{67.9} & 54.5 & 67.4 \\
Qwen3-Omni-30B-A3B-Captioner   & 66.4 & \underline{55.3} & \underline{68.2} \\
\midrule
\textbf{ACE-Cap (ours)} & \textbf{73.0} & \textbf{64.1} & \textbf{70.4} \\
\bottomrule
\end{tabular}
\caption{Caption-as-evidence accuracy (\%) on the audio-only
MMAU, MMAR, and MMSU evaluations. Qwen3.6-27B answers each
question using only the generated caption as audio-derived evidence.
All entries are reserved for results obtained under the unified
protocol. Best and second-best results among open-source models are highlighted in bold and underlined, respectively.}
\label{tab:caption-evidence-qa}
\end{table}

\begin{table}[t]
\centering
\footnotesize
\setlength{\tabcolsep}{6pt}
\begin{tabular}{@{}lc@{}}
\toprule
\textbf{Method} & \textbf{Omni-Cloze $\uparrow$} \\
\midrule
\multicolumn{2}{@{}l}{\textit{\textbf{Proprietary models}}} \\
GPT-4o Audio                         & 35.8 \\
Gemini2.0-Flash                     & 20.0 \\
Gemini2.5-Flash                     & 42.6 \\
Gemini2.5-Pro                       & 48.0 \\
Gemini3.1-Pro                       & 64.1 \\
Qwen3.5-Omni-Plus                   & 57.6 \\
\midrule
\multicolumn{2}{@{}l}{\textit{\textbf{Open-source models}}} \\
Kimi-Audio-7B                        &  2.9 \\
Step-Audio-2-mini-8B                 &  5.7  \\
Audio-Flamingo 3-7B                  &  6.3 \\
SALMONN-13B                          & 10.6 \\
MiDashengLM-7B                       & 19.5 \\
Qwen2-Audio-7B                       & 22.2 \\
Qwen2.5-Omni-7B                      & 25.8 \\
Audio-Captioner-7B                   & 53.2 \\
Qwen3-Omni-30B-A3B-Instruct          & 35.3 \\
Qwen3-Omni-30B-A3B-Captioner         & \underline{57.5}	 \\
\midrule
\textbf{ACE-Cap (ours)}              & \textbf{64.4} \\
\bottomrule
\end{tabular}
\caption{Audio-only results on the Omni-Cloze audio subset. The metric
is official overall accuracy (\%). All entries are reserved for
results obtained under the same caption-only Qwen3.6-27B evaluation
protocol.}
\label{tab:omnicloze-audio-only}
\end{table}

\section{Experiments}
\label{sec:experiments}

\begin{figure*}[t]
    \centering
    \begin{minipage}[t]{0.43\textwidth}
        \centering
        \includegraphics[width=\linewidth]{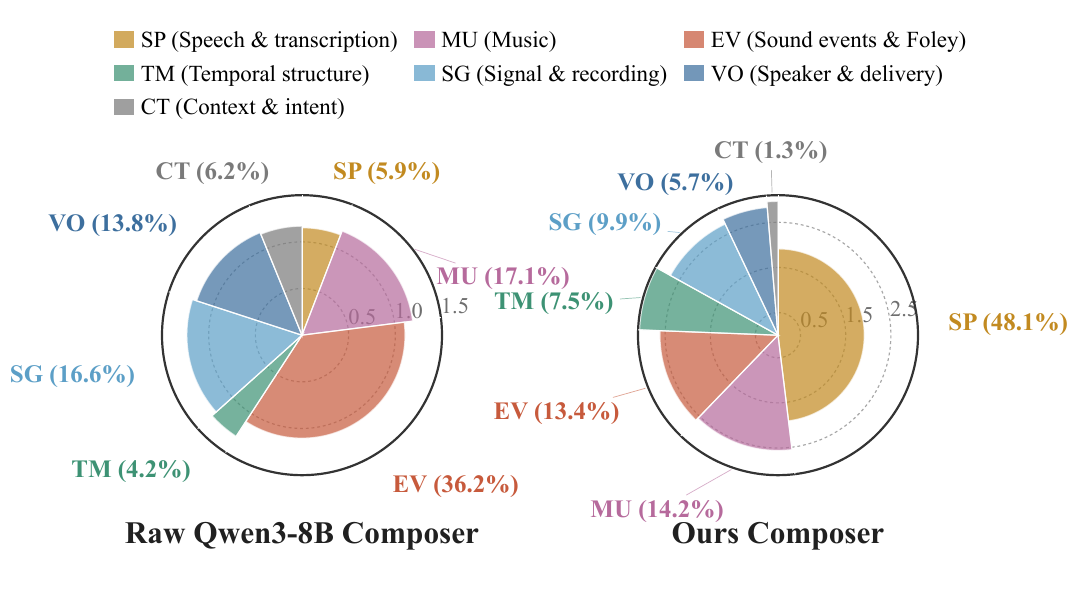}
        \par\smallskip
        \textbf{(a) Composer query behavior}
    \end{minipage}
    \hfill
    \begin{minipage}[t]{0.43\textwidth}
        \centering
        \includegraphics[width=\linewidth]{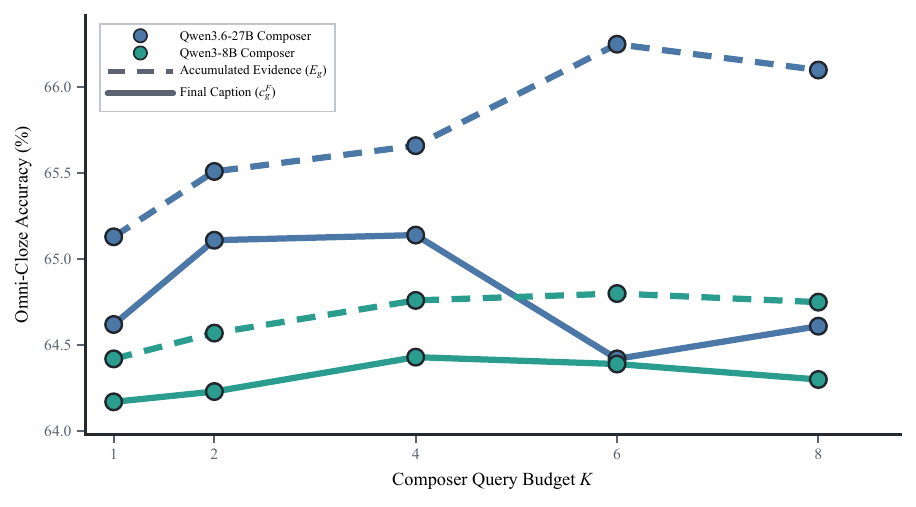}
        \par\smallskip
        \textbf{(b) Query-budget ablation}
    \end{minipage}
    \caption{Composer analysis on Omni-Cloze. (a) Query-category behavior:
    sector angle denotes the proportion of queries assigned to each category,
    while radius denotes the category-specific mean query turn. (b) Accuracy
    versus the maximum Composer query budget $K$. Dashed and solid lines
    evaluate the Accumulated Evidence $E_g$ and final caption $c_g^F$,
    respectively. Blue denotes the frozen base Qwen3.6-27B Composer, which was
    not trained with LOOP-GRPO or alternating co-evolution. Teal denotes our
    Qwen3-8B Composer, which was trained with both.}
    \label{fig:composer-analysis}
\end{figure*}

\begin{figure*}[h]
    \centering
    \includegraphics[width=0.8\textwidth]{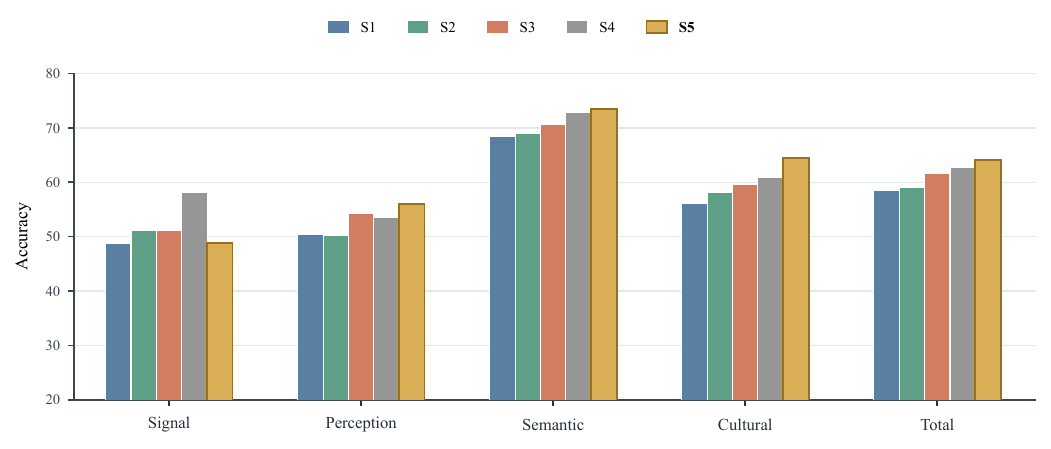}
    \caption{Ablation of the ACE-Cap alternating training pipeline on MMAR.
    S1--S5 follow the ordered training stages in
    Algorithm~\ref{alg:ace-cap}: S1 is the all-SFT initialization; S2 applies
    Captioner GRPO after S1; S3 further applies Instruct GRPO; S4 further
    applies Composer LOOP-GRPO warm-up; and S5 is the final ACE-Cap model after
    alternating Composer--Instruct co-evolution. Accuracy (\%) is reported for
    the four MMAR capability layers and the full test set. The y-axis begins
    at 20\%. The corresponding single- and mixed-modality results are
    reported in Appendix~C.1 of the supplementary document.}
    \label{fig:mmar-training-stages}
\end{figure*}

\subsection{Experimental Setup}
\label{sec:experimental-setup}

\paragraph{Training Data and Implementation Details.}
ACE-Cap is trained on 50,000 audio clips sampled from
ASID-1M~\cite{ASID}. We use Gemini 3.1 Pro-Preview%
\footnote{Official model card: \url{https://deepmind.google/models/model-cards/gemini-3-1-pro/}.}
(hereafter Gemini 3.1 Pro) to generate gold captions for all clips and
Gold-MCQs from a 3,000-pair subset. The Captioner and Instruct models
are initialized from Qwen2.5-Omni-7B~\cite{Qwen25-Omni}, while the
Composer is initialized from Qwen3-8B~\cite{Qwen3}. Further
data-construction and implementation details are provided in Appendix~A
of the supplementary document, with the complete annotation prompts in
Appendix~D.

\paragraph{Baselines and Evaluation Benchmarks.}
Following Omni-Captioner~\cite{Omni-Captioner}, we evaluate all
comparison systems as caption generators under a unified
caption-as-evidence protocol. Proprietary baselines include GPT-4o
Audio~\cite{gpt4o-audio}, Gemini 2.0 Flash~\cite{gemini}, Gemini 2.5
Flash and Gemini 2.5 Pro~\cite{gemini25}, Gemini 3.1 Pro, and
Qwen3.5-Omni-Plus~\cite{Qwen35Omni}. Open-source baselines include
Audio-Flamingo 3-7B~\cite{AudioFlamingo3}, Kimi-Audio-7B
\cite{Kimi-Audio}, Step-Audio-2-mini-8B~\cite{Step-Audio2},
SALMONN-13B~\cite{Salmonn}, and MiDashengLM-7B~\cite{Midashenglm}.
We further include Qwen2-Audio-7B~\cite{Qwen2-Audio},
Qwen2.5-Omni-7B~\cite{Qwen25-Omni}, Audio-Captioner-7B
\cite{Omni-Captioner}, and the Qwen3-Omni-30B-A3B-Instruct and
Qwen3-Omni-30B-A3B-Captioner variants~\cite{Qwen3-Omni}. Evaluation
covers Omni-Cloze~\cite{Omni-Captioner}, MMAR~\cite{MMAR},
MMAU~\cite{MMAU}, and MMSU~\cite{MMSU}, using the same frozen
Qwen3.6-27B~\cite{qwen36-27b} text-only judge. Detailed descriptions of
the baselines, benchmarks, and evaluation protocol are provided in
Appendix~B of the supplementary document.

\subsection{Main Comparison}
\label{sec:main-comparison}

\paragraph{Caption utility on audio question answering.}
Table~\ref{tab:caption-evidence-qa} compares the extent to which each
caption can substitute for the audio on the three downstream
question-answering benchmarks.
ACE-Cap achieves 73.0\%, 64.1\%, and 70.4\% on MMAU, MMAR, and MMSU,
respectively. It outperforms the strongest open-source baseline by 5.1, 8.8,
and 2.2 percentage points on the three benchmarks. Across the 18 comparisons
with proprietary models, ACE-Cap records 15 higher scores, one tie, and two
lower scores. It exceeds GPT-4o Audio and Qwen3.5-Omni-Plus on all three
benchmarks. Compared with Gemini 2.5 Pro, it leads MMAU by 3.0 points, matches
MMAR, and trails MMSU by 1.4 points. Compared with Gemini 3.1 Pro, it leads
MMAU and MMSU by 1.7 and 0.9 points, while trailing MMAR by 1.6 points. The
MMAR result therefore remains competitive with the strongest proprietary
system, while the broader pattern indicates strong caption utility across
varied audio reasoning tasks.

\paragraph{Fine-grained evidence coverage on Omni-Cloze.}
Table~\ref{tab:omnicloze-audio-only} reports the corresponding
audio-only Omni-Cloze comparison.
On Omni-Cloze, ACE-Cap achieves 64.4\%, the highest accuracy among all evaluated methods. This exceeds the strongest open-source model,
Qwen3-Omni-30B-A3B-Captioner, by 6.9 points. ACE-Cap also outperforms every
proprietary baseline, including GPT-4o Audio, Gemini 2.5 Pro,
Qwen3.5-Omni-Plus, and Gemini 3.1 Pro. The margins over these models are 28.6,
16.4, 6.8, and 0.3 points, respectively. The 0.3-point margin over Gemini 3.1
Pro is modest, so we interpret the result as the best point estimate rather
than a large separation. Under the unified caption-only evaluation protocol,
this result shows that ACE-Cap retains fine-grained evidence needed to resolve
masked acoustic details.

\subsection{Analysis \& Ablation Studies}
\label{sec:analysis}

\paragraph{Ablation of the ACE-Cap Alternating Training Pipeline.}

Following Algorithm~\ref{alg:ace-cap}, Fig.~\ref{fig:mmar-training-stages}
evaluates the successive stages of the ACE-Cap alternating training pipeline.
Total accuracy improves monotonically from 58.6\% at S1 to 64.1\% at S5.
Captioner GRPO yields a modest 0.5-point gain at S2, whereas adding Instruct
GRPO produces the largest improvement of 2.5 points at S3, together with clear
gains on the Perception and Semantic layers. Composer LOOP-GRPO warm-up further
raises the total by 1.1 points at S4, and alternating co-evolution adds another
1.4 points at S5. From S1 to S5, Perception, Semantic, and Cultural accuracy
increase by 5.44, 5.09, and 8.51 points, respectively, indicating that the
later training stages mainly improve higher-level evidence acquisition and
integration. Signal accuracy is less stable: it peaks at S4 and returns to its
S1 value at S5, partly reflecting the small size of this subset ($n=43$).
Thus, the overall gain is not caused by a uniform increase across all layers.
Each variant retains the preceding training stages, so adjacent comparisons
quantify the incremental effect of the newly introduced stage within the
ordered pipeline rather than an isolated component-removal effect. Results for
speech, sound, music, and their mixed-modality subsets are provided in
Appendix~C.1 of the supplementary document.

\paragraph{Active Evidence Acquisition}
\label{sec:active-evidence-analysis}

Fig.~\ref{fig:composer-analysis}(a) shows that the Raw Composer
distributes its queries more broadly, with a strong preference for sound
events (EV, 36.2\%), and asks most questions near the first turn. In contrast,
Ours Composer focuses more on speech and transcription (SP, 48.1\%) while
placing temporal, speaker-related, and contextual queries in later turns.
This shift indicates a more focused and structured evidence acquisition
strategy.

\paragraph{Query-Budget Ablation.}

Fig.~\ref{fig:composer-analysis}(b) evaluates how the maximum query budget
affects evidence acquisition and preservation. We measure preservation using
$\Delta_{\mathrm{ret}}(K)=
\operatorname{Acc}(E_g)-\operatorname{Acc}(c_g^F)$. The two policies differ
in both backbone and training, so this comparison does not isolate the causal
effect of LOOP-GRPO. We therefore treat the retention gap as a diagnostic
rather than a controlled training ablation.

For the frozen Qwen3.6-27B Composer, $\Delta_{\mathrm{ret}}$ is
0.40--0.52 percentage points (pp) for $K\leq4$, but increases to 1.83 and
1.49 pp at $K=6$ and $K=8$, respectively. Thus, the higher accumulated-evidence
accuracy at larger budgets is not preserved by the final caption. In contrast,
our Qwen3-8B Composer maintains a narrower gap of 0.25--0.45 pp across all
budgets. This pattern is consistent with the objective of LOOP-GRPO's
turn-level, stopping, and final-caption credits, which are designed to improve
multi-turn credit assignment and evidence-preserving synthesis. For our
Composer, $K=2$--$4$ provides a favorable accuracy--cost trade-off: budgets
above four increase interaction cost without improving final-caption accuracy.
Complete results for all evaluated configurations are reported in
Appendix~C.2 of the supplementary document.

\section{Conclusion}

We presented ACE-Cap, an active framework that reformulates long-paragraph,
fine-grained audio captioning as iterative evidence acquisition. A text-only
Composer identifies missing information, queries an audio-conditioned
Instruct model, and synthesizes the accumulated evidence, while LOOP-GRPO,
role-wise warm-up, and alternating optimization train effective questioning,
stopping, and evidence-preserving generation. Across the evaluated audio
benchmarks, ACE-Cap consistently outperforms strong open-source baselines and
remains competitive with proprietary systems, supporting active interaction
as an effective strategy for detailed audio understanding. The current study
is limited to audio-only inputs and a fixed caption-as-evidence evaluation
protocol; future work will consider audio-visual evidence, broader evaluation,
and adaptive interaction budgets.

\bibliography{aaai2027}

\end{document}